\documentclass[12pt]{article}
\usepackage{graphicx}

\newcommand\pubnumber{Article 13 in eConf C1304143}
\newcommand\pubdate{\today}

\def\Title#1{\begin{center} {\Large #1 } \end{center}}
\def\Author#1{\begin{center}{ \sc #1} \end{center}}
\def\Address#1{\begin{center}{ \it #1} \end{center}}

\newcommand\pubblock{\rightline{\begin{tabular}{l} \pubnumber\\
         \pubdate  \end{tabular}}}
\newenvironment{Abstract}{\begin{quotation}  }{\end{quotation}}
\newenvironment{Presented}{\begin{quotation} \begin{center} 
             PRESENTED AT\end{center}\bigskip 
      \begin{center}\begin{large}}{\end{large}\end{center} \end{quotation}}
\def\Acknowledgements{\bigskip  \bigskip \begin{center} \begin{large}
             \bf ACKNOWLEDGEMENTS \end{large}\end{center}}

\def\beq{\begin{equation}}
\def\eeq#1{\label{#1}\end{equation}}
\def\eeqn{\end{equation}}

\def\beqa{\begin{eqnarray}}
\def\eeqa#1{\label{#1}\end{eqnarray}}
\def\eeqan{\end{eqnarray}}

\let\bar=\overbar

\def\Dslash{\not{\hbox{\kern-4pt $D$}}}
\def\dslash{\not{\hbox{\kern-2pt $\del$}}}

\def\msb{{\bar{\ssstyle M \kern -1pt S}}}

\begin{document}
\begin{titlepage}
\pubblock

\vfill
\Title{Novel distance indicator for Gamma-Ray Bursts associated with Supernovae}
\vfill
\Author{G. B. Pisani$^{1,3}$,  L. Izzo$^{1,2}$, R. Ruffini$^{1,2}$, C. L. Bianco$^{1,2}$, M. Muccino$^{1}$, A. V. Penacchioni$^{1,3}$, J. A. Rueda$^{1,2}$, Y. Wang$^{1}$}
\Address{$^{1}$Dipartimento di Fisica, Sapienza Universit\'a di Roma and ICRA, Piazzale Aldo Moro 5, I-00185 Roma, Italy,\\
$^{2}$ICRANet, Piazza della Repubblica 10, I-65122 Pescara, Italy,\\
$^{3}$Universit\'e de Nice Sophia Antipolis, Nice, CEDEX 2, Grand Chateau Parc Valrose.}

\vfill

\begin{Abstract}
It has been proposed that the temporal coincidence of a gamma-ray burst (GRB) and a type Ib/c supernova (SN) can be explained with the concept of induced gravitational collapse (IGC), induced by the matter ejected from an SN Ib/c accreting onto a neutron star (NS). We found a standard luminosity light curve behavior in the late-time X-ray emission of this subclass of GRBs. We interpret this as the result of a common physical mechanism in this particular phase of the X-ray emission, possibly related to the creation of the NS from the SN process. Moreover, this scaling law could be a fundamental tool for estimating the redshift of GRBs that belong to this subclass of events.
\end{Abstract}
\vfill
\begin{Presented}
Huntsville in Nashville GRB Symposium\\
Nashville, TN, USA,  April 14--18, 2013
\end{Presented}
\vfill
\end{titlepage}
\def\thefootnote{\fnsymbol{footnote}}
\setcounter{footnote}{0}

\section{Introduction}

Recently, in \cite{Ruffini2001c,Ruffini2007b}, it was proposed that the temporal coincidence of some gamma-ray bursts (GRBs) and a type Ib/c supernovae (SNe) can be explained with the concept of induced gravitational collapse (IGC) of a neutron star (NS) to a black hole (BH) induced by accretion of matter ejected by the SN Ib/c. More recently, this concept has been extended in \cite{Rueda2012}, including a precise description of the progenitor system of such GRB-SN systems.

The main new result presented here is that the IGC GRB-SN class shows a standard late X-ray luminosity light curve in the common energy range $0.3\,$--$\,10$ keV \cite{RuffiniMG13,Pisani2013}.

The prototype is GRB 090618 \cite{TEXAS,Izzo2012,Izzo2012b} at redshift $z=0.54$, where four different emission episodes have been identified:
\begin{itemize}
\item Episode 1, corresponding to the SN onset, has been observed to have thermal as well as power-law emission. The thermal emission changes in time following a precise power-law behavior \cite{Izzo2012,Penacchioni2012,Ana2013};
\item Episode 2 follows and in the IGC model corresponds to the GRB emission coincident with the BH formation. The characteristic parameters of the GRB, including baryon load, the Lorentz factor, and the nature of the circumburst medium (CBM), have been computed \cite{Izzo2012,Penacchioni2012,Ana2013};
\item Episode 3 is characterized in the X-ray light curve by a shallow phase (a plateau) followed by a final steeper decay. Typically, it is observed in the range $10^2\,$--$\,10^6$ s after the GRB trigger;
\item Episode 4 occurs after a time of about ten days in the cosmological rest-frame, corresponding to the SN emission due to the Ni decay \cite{Arnett}. This emission is clearly observed in GRB 090618 during the late optical GRB afterglow emission.
\end{itemize}

\section{Observations}

Here we analyze the X-ray emission of a sample of eight GRBs with $E_{iso} \geq 10^{52}$ erg that satisfy at least one of the following three requirements: there is a double emission episode in the prompt emission: Episode 1, with a decaying thermal feature, and Episode 2, a canonical GRB, as in GRB 090618 \cite{Izzo2012}, GRB 101023 \cite{Penacchioni2012}, and in GRB 110709B \cite{Ana2013}; there is a shallow phase followed by a final steeper decay in the X-ray light curve: Episode 3; an SN is detected after about ten days from the GRB trigger in the rest-frame: Episode 4.

\begin{table}[t]
\begin{center}
\begin{tabular}{l c c}
\hline\hline
GRB  & $z$ & $E_{iso} (erg)$ \\
\hline 
GRB 060729 & $0.54$ & $1.6 \times 10^{52}$ \\
GRB 061007 & $1.261$ & $1.0 \times 10^{54}$ \\
GRB 080319B & $0.937$ & $1.3 \times 10^{54}$ \\
GRB 090618 & $0.54$ & $2.9 \times 10^{53}$ \\
GRB 091127 & $0.49$ & $1.1 \times 10^{52}$ \\
GRB 111228 & $0.713$ & $2.4 \times 10^{52}$ \\
\hline
GRB 101023 & $0.9^*$ & $1.8 \times 10^{53}$ \\
GRB 110709B & $0.75^*$ & $1.7 \times 10^{53}$ \\
\hline\\
\end{tabular}
\caption{GRB sample considered in this work. The redshifts of GRB 101023 and GRB 110709B, which are marked with an asterisk, were deduced theoretically by using the method outlined in \cite{Penacchioni2012} and the corresponding isotropic energy computed by assuming these redshifts. More details on the sources of this sample are in \cite{Pisani2013}.}
\label{table1}
\end{center}
\end{table}

\begin{figure}
\centering
\includegraphics[width=0.78\hsize,clip]{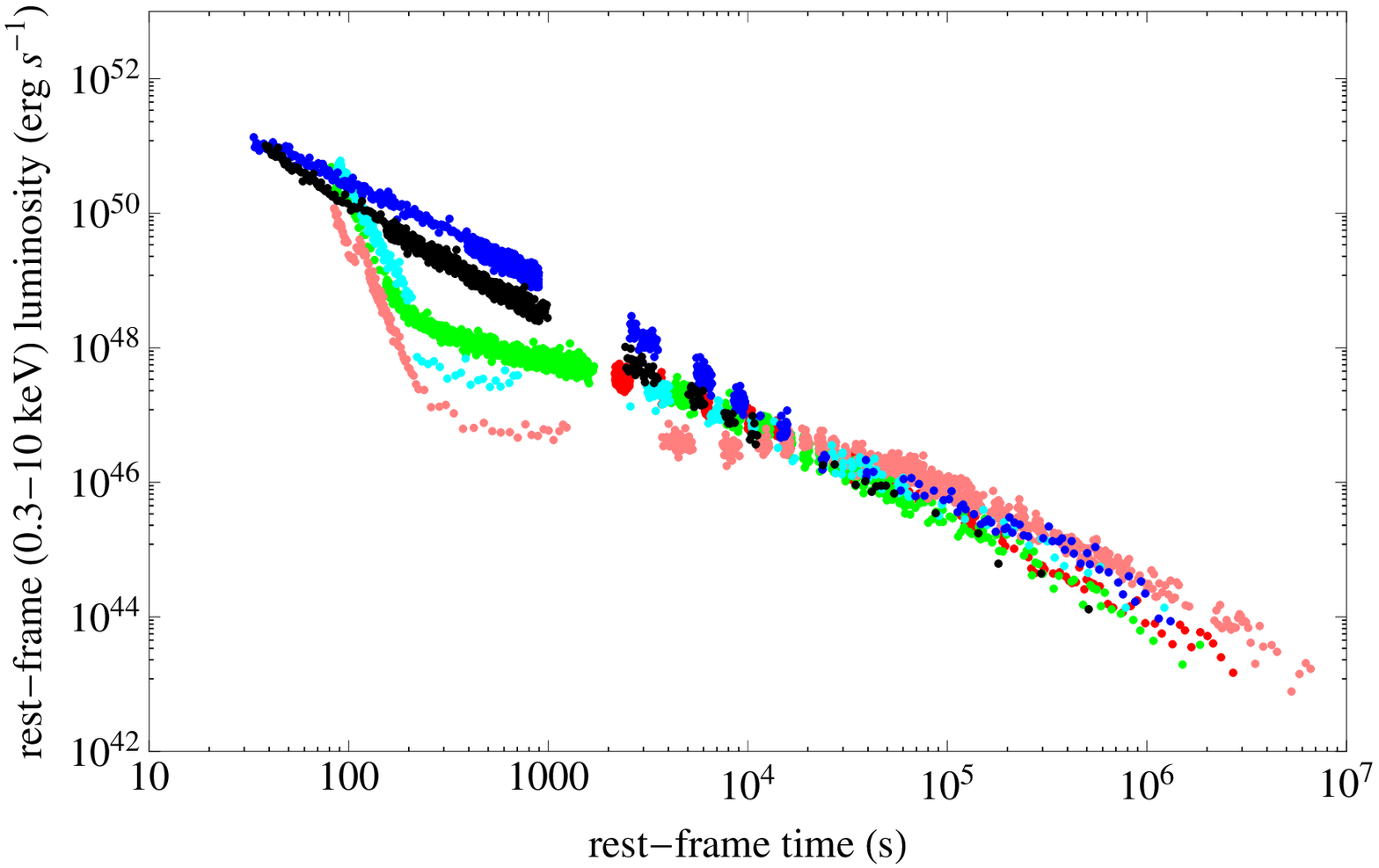}
\caption{X-ray luminosity light curves of the six GRBs with measured redshift in the $0.3\,$--$\,10$ keV rest-frame energy range: in pink GRB 060729, $z=0.54$; black GRB 061007, $z=1.261$; blue GRB 080319B, $z=0.937$; green GRB 090618, $z=0.54$, red GRB 091127, $z=0.49$, and in cyan GRB 111228, $z=0.713$.}\label{fig:sample}
\end{figure}

We found eight GRBs that satisfy our requirements (see Table \ref{table1}). For further details on the sample, see \cite{Pisani2013}. We focused the analysis of all available XRT data of these sources. Characteristically, XRT follow-up starts only about 100 seconds after the BAT trigger (typical repointing time of Swift after the BAT trigger). Because the behavior was similar in all sources, we compared the analyzed XRT luminosity light curve for the six GRBs with measured redshifts in the common rest-frame energy range $0.3\,$--$\,10$ keV. As a first step we converted the observed XRT flux to the one in the $0.3\,$--$\,10$ keV rest-frame energy range. In the detector frame, the $0.3\,$--$\,10$ keV rest-frame energy range becomes $[0.3/(1+z)]\,$--$\,[10/(1+z)]$ keV, where $z$ is the redshift of the GRB. We assumed a simple power-law function as the best fit for the spectral energy distribution of the XRT data\footnote{http://www.swift.ac.uk/}. For details about this computation, see \cite{Pisani2013}. 

The X-ray luminosity light curves of the six GRBs with measured redshifts in the $0.3$--$10$ keV rest-frame energy band are plotted in Fig. \ref{fig:sample}. What is most striking is that these six GRBs, with redshifts in the range $0.49\,$--$\,1.261$, show a remarkably common behavior of the late X-ray afterglow luminosity light curves (Episode 3), despite their very different prompt emissions (Episode 1 and 2) and energetics spanning more than two orders of magnitude. The common behavior starts between $10^4\,$--$\,10^5$ s after the trigger and continues until the emission falls below the XRT threshold.

\section{Interpretation}

This standard behavior of Episode 3 represents a strong evidence of very low or even absent beaming in this particular phase of the X-ray afterglow emission process. We have proposed that this late-time X-ray emission in Episode 3 is related to the process of the SN explosion within the IGC scenario, possibly emitted by the newly born NS, and not by the GRB itself \cite{Negreiros2012}.

\begin{figure}
\centering
\includegraphics[width=0.49\hsize,clip]{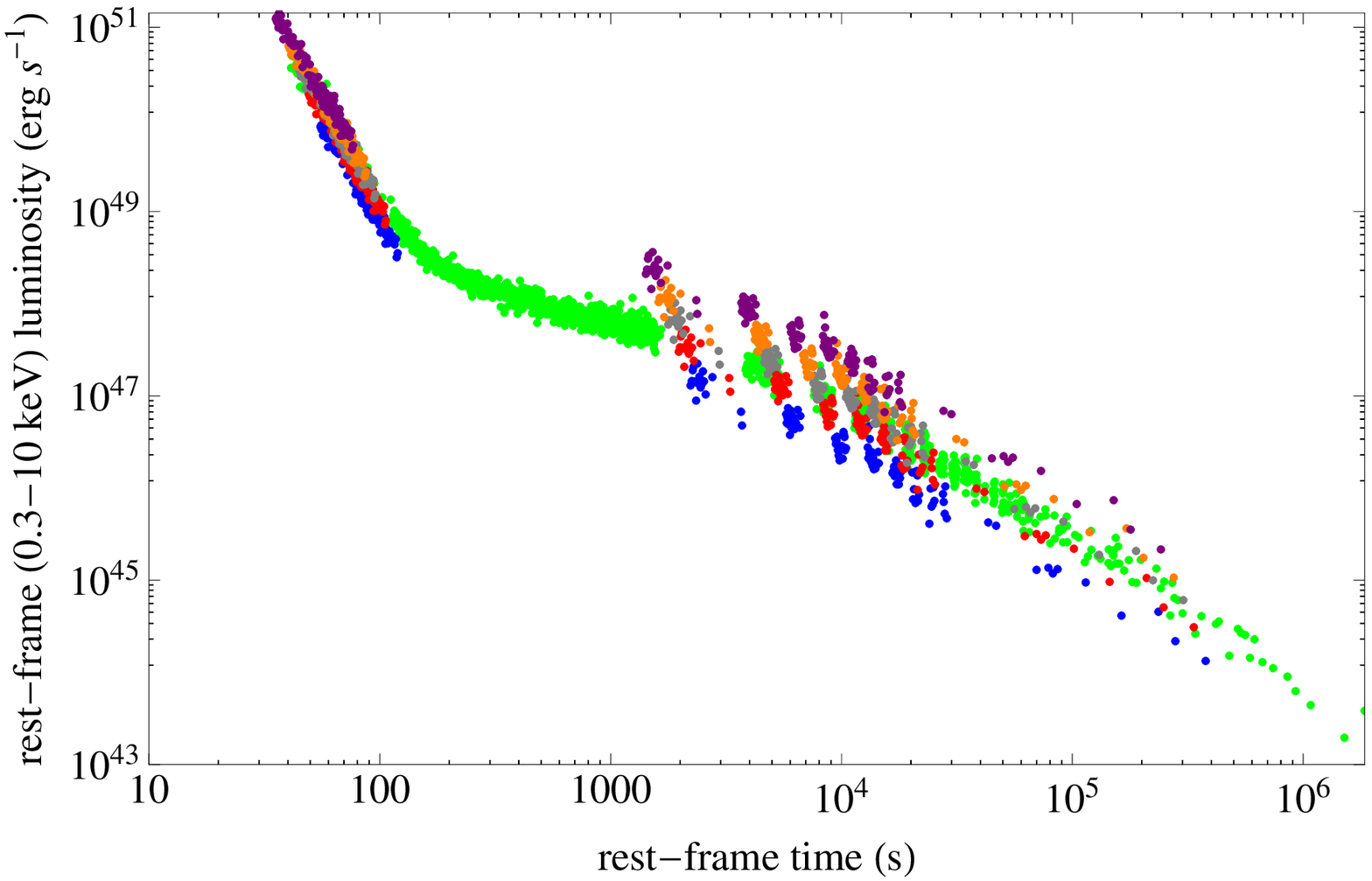}
\includegraphics[width=0.49\hsize,clip]{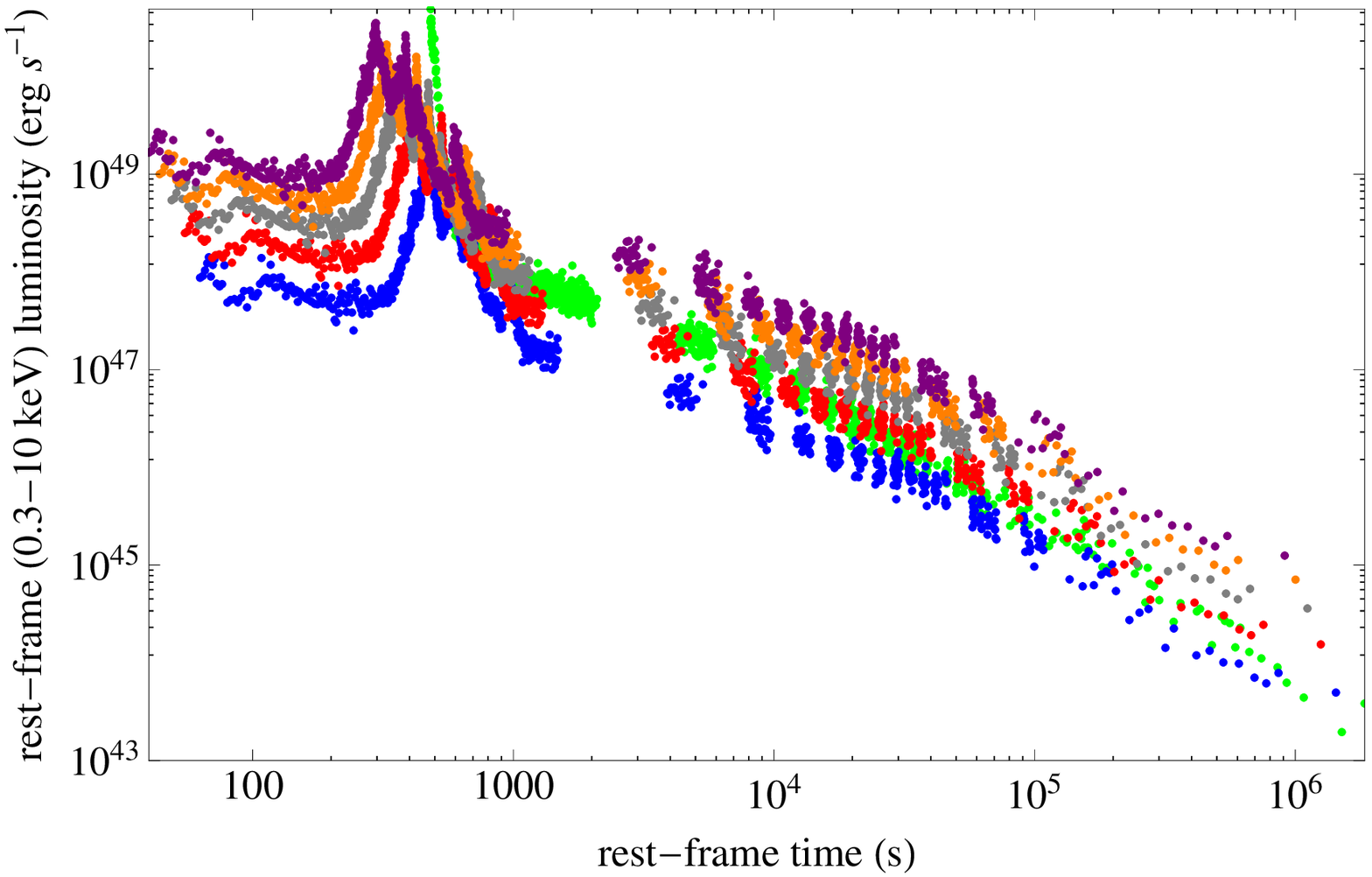}
\caption{In green we show the rest-frame X-ray luminosity light curve of GRB 090618 in the $0.3\,$--$\,10$ keV energy range in comparison with the one of GRB 101023 (left) and GRB 110709B (right), computed for different hypothetical redshifts: respectively, from blue to purple: $z=0.6, 0.8, 1.0, 1.2, 1.5$ (left) and $z=0.4, 0.6, 0.8, 1.0, 1.2$ (right). The overlapping at late time of the two X-ray luminosity light curves is obtained for a redshift of $z=0.9$ (left) and $z=0.75$ (right). For further details see \cite{Penacchioni2012,Ana2013}.}
\label{fig:101023}
\end{figure}

This scaling law, when confirmed in sources with Episode 1 plus Episode 2 emissions, offers a powerful tool for estimating the redshift of GRBs that belong to this subclass of events. As an example, Fig.~\ref{fig:101023} plots the rest-frame X-ray luminosity (0.3 - 10 keV) light curve of GRB 090618 (considered the prototype of the common behavior shown in Fig. \ref{fig:sample}) with the rest-frame X-ray luminosity light curves of GRB 110709B estimated for selected values of its redshifts $z=0.4, 0.6, 0.8, 1.0$, and $1.2$, and similarly the corresponding analysis for GRB 101023 for redshifts $z=0.6, 0.8, 1.0, 1.2$, and $1.5$. This shows that GRB 101023 should have been located at $z \sim 0.9$ and GRB 110709B at $z \sim 0.75$. These redshift estimates are within the range expected using the Amati relation as shown in \cite{Penacchioni2012,Ana2013}. This is an important independent validity confirmation for this new redshift estimator we are proposing for the family of IGC GRB-SN systems. We stress, however, that the redshift was determined assuming the validity of the standard $\Lambda$CDM cosmological model for sources with redshift in the range $z=0.49\,$--$\,1.216$. We are currently testing the validity of this assumption for sources at higher cosmological redshifts.

We are currently testing the predictive power of our results on three different observational scenarios for sources of the IGC GRB-SN subclass:
\begin{itemize}
\item GRBs at high redshift. We are able to predict the existence of an SN in these systems, which is expected to emerge after $t \sim 10 \, (1+z)$ days, the canonical time sequence of an SN explosion. This offers a new challenge to detect SNe at high redshift, e.g., by observing radio emission \cite{Ana2013};
\item for GRBs with $z\leq1$  we can indicate in advance from the X-ray luminosity light curve observed by XRT the expected time for the observations of an SN and alert direct observations from ground- and space-based telescopes
\item as we showed here, we can infer the redshift of GRBs in the same way we did for GRB 110709B and GRB 101023A.
\end{itemize}

We are currently expanding the sample to increase the statistical validity of our approach and its cosmological implications.

\Acknowledgements
We are grateful to the \textit{Swift} team for the support. This work made use of data supplied by the UK \textit{Swift} Data Center at the University of Leicester. G.B. Pisani and A.V. Penacchioni are supported by the Erasmus Mundus Joint Doctorate Program by Grant Numbers 2011-1640 and 2010-1816, respectively, from the EACEA of the European Commission.

\end{document}